\documentstyle[preprint,prd,aps]{revtex}
\advance\textheight by \baselineskip
\begin{document}
\draft

\newcommand{\Det}{{\rm Det}}
\newcommand{\Tr}{{\rm Tr}}
\newcommand{\Log}{{\rm Log}}
\newcommand{\Dslash}{{D \hskip -8pt /}}

\preprintstytrue
\preprint{\hbox to\textwidth{UCLA/96/TEP/1 \hfill
    hep-th/9601048 \hfill UCONN-96-1}}

\title{Temperature Expansions for Magnetic Systems}

\author{Daniel Cangemi\cite{emaila}}

\address{ Department of Physics, University of California,
  Los Angeles, CA 90095-1547 }

\author{Gerald Dunne\cite{emailb}}

\address{ Department of Physics, University of Connecticut, Storrs, CT
  06269-3046 }


\maketitle

\begin{abstract}
  We derive finite temperature expansions for relativistic fermion
  systems in the presence of background magnetic fields, and with
  nonzero chemical potential. We use the imaginary-time formalism for
  the finite temperature effects, the proper-time method for the
  background field effects, and zeta function regularization for
  developing the expansions. We emphasize the essential difference
  between even and odd dimensions, focusing on $2+1$ and $3+1$
  dimensions. We concentrate on the high temperature limit, but we
  also discuss the $T=0$ limit with nonzero chemical potential.
\end{abstract}



\section{Introduction}

The study of fermion systems in the presence of external
electromagnetic fields has applications in diverse areas of physics,
including astrophysics, solid state, condensed matter, plasma and
particle physics. Indeed, astrophysical considerations led to the
first systematic study of relativistic noninteracting fermion systems
\cite{chandra}. The existence of very high intensity magnetic fields
in gravitationally collapsed objects motivated further investigations
of the energy-momentum tensor and equation of state for a degenerate
electron gas in a strong uniform magnetic field \cite{chiu}. More
recently, this problem has been addressed in the framework of finite
temperature quantum field theory \cite{elmfors,zeitlin,danielsson},
using the fact that many of the thermodynamic properties may be
derived from the corresponding finite temperature effective action.

In this paper we present a pedagogical discussion of the finite temperature
($T$) and finite chemical potential ($\mu$) effective action for fermions in an
external static magnetic field. Our aim is to review and unify the
imaginary-time approach \cite{dolan,shuryak,kapusta}, developed initially for
finite temperature {\it free} systems , with the Fock-Schwinger proper-time
approach \cite{schwinger,salam,blau} developed for systems interacting with
external electromagnetic fields but at {\it zero} temperature. Formally, these
two approaches fit together beautifully for the case of static external
magnetic fields; in fact, the finite $T$ and finite $\mu$ effects are {\it
completely separate} from the computation required to compute the effects of
the external static magnetic field. Thus, knowledge of the zero $T$ and zero
$\mu$ proper-time effective action is completely sufficient to write down a
corresponding expression at nonzero temperature and nonzero chemical potential.
However, this expression for the effective action is formal and the separation
between $T$ and $\mu$ effects and those of the external field may become
blurred when one tries to make various approximate asymptotic expansions, such
as for high or low temperature. Such expansions are necessarily complicated due
to the proliferation of energy scales: thermal energy $kT$, fermion mass $m$,
chemical potential $\mu$, cyclotron energy $B/m$, and also possible
momentum scales associated with spatial variations of the external field. We
concentrate mainly on the high temperature limit, in which $T$ (we
use units in which Boltzmann's constant $k=1$) is the dominant energy scale.
However, we also present a simple approach to the other extreme: the $T=0$
limit. This complements Refs. \cite{elmfors,zeitlin,danielsson}, which
have focussed on the low $T$ limit, and which have primarily used the real-time
formalism for discussing the finite temperature effects.

Throughout our analysis, we treat both $2+1$ dimensional and $3+1$ dimensional
theories. This is motivated by the known profound differences between $2+1$ and
$3+1$ dimensions (and in general between odd and even dimensional space-times)
for free fermion systems at finite $T$ and $\mu$ \cite{actor,weldon}, and for
zero $T$ fermions in external fields \cite{blau}. We find that a consistent
treatment of both cases requires careful use of zeta function regularization,
which has been used previously for free systems \cite{actor,weldon} (and for
systems with constant external $A_0$ \cite{actor2}). For the magnetic
backgrounds we also find that the $2+1$ and $3+1$ cases involve very different
expansions at high temperature. The high temperature behavior of ${\rm
QED}_{2+1}$ is also of interest for studying questions of spontaneous symmetry
breaking \cite{roman,pisarski,appelquist,poly,miransky}.

In Section II we review the structure of the zero temperature effective action
with nonzero chemical potential for $2+1$ and $3+1$ dimensional fermions in
external static magnetic fields. Finite temperature is introduced in Section
III using the imaginary-time formalism. In Section IV this is combined with the
proper-time formalism to provide a general formal expression for the finite $T$
and $\mu$ effective action. In Section V we apply this to the high temperature
limit of the free fermionic theories, and in Section VI to the high temperature
limit of fermions in a static magnetic field. The zero temperature limit is
examined in Section VII, and we conclude in Section VIII with
some comments regarding possible further extensions of this approach.

\section{Effective Action for Magnetic Systems}

The basic object of interest in this paper is the effective action
\begin{equation}
  i S_{\rm eff}=\Log\, \Det \left(i\Dslash -m -\mu \gamma^0\right)
\label{eff}
\end{equation}
where $\Dslash = D_\nu
\gamma^\nu=\left(\partial_\nu+ieA_\nu\right)\gamma^\nu$, and we choose
Minkowski gamma matrices $\gamma^\nu$ satisfying
\begin{equation}
  \{ \gamma^\nu , \gamma^\sigma \}= -2 g^{\nu\sigma}=2 \, {\rm
    diag}(1,-1,-1,\dots, -1)
\label{gamma}
\end{equation}
The term $\mu \gamma^0$ in the Dirac operator in~(\ref{eff}) reflects
the presence of a chemical potential $\mu$, corresponding to a term
$-\mu\psi^\dagger\psi$ in the Lagrangian.

In $2+1$ dimensions, the irreducible gamma matrices may be chosen to
be
\begin{equation}
  \gamma^0=\sigma^3\hskip 2cm \gamma^1=i\sigma^1 \hskip 2cm
  \gamma^2=i\sigma^2
\label{3dgamma}
\end{equation}
where the $\sigma^i$ are the $2\times 2$ Pauli matrices. Note that an
alternative choice, $\gamma^0=-\sigma^3$, $\gamma^1=-i\sigma^1$,
$\gamma^2=-i\sigma^2$, corresponds to changing the sign of the mass,
$m\to-m$, in the Dirac operator appearing in the effective
action~(\ref{eff}). The system with effective action~(\ref{eff}) is
not parity invariant since a fermion mass term breaks parity in $2+1$
dimensions \cite{roman,pisarski,appelquist,poly}.  However, a parity
invariant model may be constructed by considering two species of
fermions, one of mass $m$ and the other of mass $-m$ (see Footnote 11 in Ref.
\cite{roman}). This may be
achieved by choosing a reducible set of $4\times 4$ gamma matrices
\begin{equation}
  \Gamma^\nu=\left(\matrix{\gamma^\nu &0\cr 0&-\gamma^\nu\cr}\right)
\label{red}
\end{equation}
in which case the Dirac operator is block diagonal:
\begin{equation}
  iD_\nu \Gamma^\nu -m -\mu \Gamma^0=\left(\matrix{i\Dslash -m -\mu
      \gamma^0 &0\cr 0& -i\Dslash -m +\mu \gamma^0\cr}\right)
\label{red-dirac}
\end{equation}
Thus, the effective action for this parity invariant system may be
written as
\begin{eqnarray}
  i S_{\rm eff}&=&\Log\,\Det_{2+1} \left(iD_\nu\Gamma^\nu -
    m-\mu\Gamma^0\right) \nonumber\\ &=&\Log \, \Det_{2+1}
  \left(-\left(iD_0-\mu\right)^2+m^2 +
    \left(\vec{D}\cdot\vec{\gamma}\right)^2\right)
\label{eff-2}
\end{eqnarray}

In $3+1$ dimensions parity symmetry is not an issue, but the effective
action (\ref{eff}), which involves the first-order Dirac operator, may
still be written in the same form as~(\ref{eff-2}), which involves a
second order operator. We use the fact that there exists an additional
gamma matrix $\gamma^5$ satisfying $\{\gamma^\nu, \gamma^5\}=0$ and
$(\gamma^5)^2={\bf 1}$.  Then
\begin{equation}
  \Det_{3+1} \left(i\Dslash -m -\mu \gamma^0\right) =
  \Det_{3+1}\left(\gamma^5\left(i\Dslash -m -\mu
      \gamma^0\right)\gamma^5\right) = \Det_{3+1}\left(-i\Dslash -m +\mu
    \gamma^0\right)
\label{4d-det}
\end{equation}
Therefore, the effective action~(\ref{eff}) may be expressed as
\begin{eqnarray}
  i S_{\rm eff}&=&{1\over 2} \Log \, \Det_{3+1} \left[\left(i\Dslash
      -m -\mu \gamma^0\right)\left(-i\Dslash -m +\mu
      \gamma^0\right)\right]\nonumber\\ &=&{1\over 2}\Log \,
  \Det_{3+1} \left(-\left(iD_0-\mu\right)^2+m^2+
    \left(\vec{D}\cdot\vec{\gamma}\right)^2\right)
\label{eff-3}
\end{eqnarray}

Note that the spatial operator
$\left(\vec{D}\cdot\vec{\gamma}\right)^2$ which appears
in~(\ref{eff-2}) and in~(\ref{eff-3}) is a positive operator. In $2+1$
dimensions it reduces to
\begin{equation}
  \left(\vec{D}\cdot\vec{\gamma}\right)^2 = -\vec{D}^2+eB\gamma^0 =
  -\left(\matrix{D_-D_+&0\cr 0&D_+D_-}\right)
\label{3d-spatial}
\end{equation}
where the magnetic field is $B=F_{12}=\partial_1 A_2-\partial_2A_1$, and
\begin{equation}
  D_\pm = D_1\pm i D_2
\label{dpm}
\end{equation}
In $3+1$ dimensions the operator
$\left(\vec{D}\cdot\vec{\gamma}\right)^2$ reduces to
\begin{equation}
  \left(\vec{D}\cdot\vec{\gamma}\right)^2=-\vec{D}^2+ie{1\over
    4}[\gamma^i, \gamma^j] F_{ij}
\label{4d-spatial}
\end{equation}
If we choose the external magnetic field to be directed along the
$x^3$ direction and to be independent of $x^3$, then (with a suitable choice of
gamma matrices) this may be simplified further to
{\renewcommand{\arraystretch}{.5}
\begin{equation}
  \left( \vec{D} \cdot \vec{\gamma} \right)^2 = - \partial_3^2 -
  \left(
  \begin{array}{cccc} D_-D_+ & 0 & 0 & 0 \\ 0 & D_+D_- & 0 & 0 \\ 0 & 0
    & D_-D_+ & 0 \\ 0 & 0 & 0 & D_+D_- \end{array} \right)
\label{4d-spatial-2}
\end{equation}
} where $D_\pm$ are as defined for the $2+1$ dimensional system
in~(\ref{dpm}).

Thus, in each case, the spectrum of the spatial operator
$\left(\vec{D}\cdot\vec{\gamma}\right)^2$ is determined by the
spectrum of the 2-dimensional Schr\"odinger-like operators $D_\pm
D_\mp$. The operator $\left(\vec{D}\cdot\vec{\gamma}\right)^2$ is
effectively diagonal and we may write
\begin{equation}
  m^2+\left(\vec{D}\cdot\vec{\gamma}\right)^2\equiv {\cal E}^2
\label{esquare}
\end{equation}
For static magnetic backgrounds $A_0=0$ and we can replace
$(iD_0-\mu)$ in (\ref{eff-2}) or~(\ref{eff-3}) by $\omega-\mu$, where
$\omega$ is an energy eigenvalue. Therefore
\begin{equation}
  iS_{\rm eff}= \int {d\omega\over 2\pi} \Tr\,\Log
  \left[-\left(\omega-\mu\right)^2+{\cal E}^2\right]
\label{eff-4}
\end{equation}
Here, the trace operation $\Tr$ is understood to mean a summation over
the eigenvalues ${\cal E}^2$ of both the positive operators $m^2 -
D_+D_-$ and $m^2 - D_-D_+$ in $2+1$ dimensions, resp. $m^2
-\partial_3^2 - D_+D_-$ and $m^2 - \partial_3^2 - D_-D_+$ in $3+1$
dimensions.

\section{Finite Temperature Formulation}

It is clear from~(\ref{eff-4}) that the effects of the external static
magnetic field are contained solely within ${\cal E}^2$, and are
clearly separated from the chemical potential $\mu$ and the energy
trace over $\omega$. We can therefore pass to a finite temperature
formulation just as in the free case \cite{dolan,shuryak,kapusta}, by replacing
the energy integration with a discrete summation:
\begin{eqnarray}
  \int {d\omega\over 2\pi} \hskip 1cm &\to& \hskip 1cm {i\over \beta}
  \sum_{n=-\infty}^\infty\nonumber\\ \omega \hskip 1cm &\to& \hskip
  1cm \omega_n={2\pi i\over \beta}\left(n+{1\over 2}\right)
\label{finite}
\end{eqnarray}
Here $\beta=1/T$, where $T$ is the temperature and Boltzmann's
constant $k$ has been absorbed into $T$. The transition to finite
temperature would not be so straightforward if there were external
{\it electric} fields, but here we consider only external static
magnetic fields. Also note that in the zero temperature
expression~(\ref{eff-4}) it looks as though the dependence on the
chemical potential $\mu$
may be formally eliminated through a naive shift of the integration variable
$\omega$. However, such a shift would violate the boundary conditions used to
compute the trace, and a proper treatment at zero temperature leads to the
appearance of non-analytic behavior in $\mu$, corresponding to sharp cut-offs
in
the energy spectrum \cite{shuryak,chodos}. At finite temperature, these sharp
cut-offs
are smoothed out, and the dependence on $\mu$ is correspondingly smooth, as we
shall see below.

The effective action~(\ref{eff-4}) may now be expressed as
\begin{eqnarray}
  S_{\rm eff}&=&{1\over \beta}\sum_{n=-\infty}^\infty \Tr\,\Log
  \left[-\Bigl(\omega_n +(|{\cal E}|-\mu)\Bigr) \Bigl(\omega_n
    -(|{\cal E}|+\mu)\Bigr)\right]\nonumber\\ &=&{1\over \beta}
  \Tr\,\Log \prod_{n=0}^\infty\left[\left( (|{\cal
        E}|-\mu)^2+(\pi(2n+1)/\beta)^2\right)\left( (|{\cal
        E}|+\mu)^2+(\pi(2n+1)/\beta)^2\right)\right]\nonumber\\
  &=&{1\over \beta} \Tr\,\Log \left[ \cosh {\beta\over 2}\left(|{\cal
        E}|-\mu\right) \cosh {\beta\over 2}\left(|{\cal E}|+\mu\right)
  \right]
\label{eff-5}
\end{eqnarray}
where in the last step we have used the infinite product
representation of the $\cosh$ function
\begin{equation}
  \cosh(x)=\prod_{n=0}^\infty \left(1+{4 x^2\over \pi^2
      (2n+1)^2}\right)
\end{equation}
and we have dropped an infinite contribution that is independent of
${\cal E}^2$.

It is a simple matter to re-write~(\ref{eff-5}) as (dropping again an
irrelevant constant)
\begin{equation}
  S_{\rm eff}= \Tr\left[|{\cal E}|+ {1\over \beta}\Log
    \left(1+e^{-\beta(|{\cal E}|-\mu)}\right) + {1\over \beta}\Log
    \left(1+e^{-\beta(|{\cal E}|+\mu)}\right)\right]
\label{eff-6}
\end{equation}
This expression for the effective action generalizes an analogous expression
(with $\mu=0$) derived in \cite{dolan} for the free case. The only effect of
the chemical potential is to shift the `energy' eigenvalue $|{\cal E}|$ by $\mp
\mu$, which corresponds to a shift in the threshold energies for particles and
antiparticles. The only effect of the external static magnetic field
is to modify the spectrum of the `energy' eigenvalue $|{\cal E}|$ from the free
spectrum to a spectrum involving dependence on the external $B$ field.
Therefore, at least in principle, we can now compute the effective action for
any external static magnetic field for which we know the spectrum of the
operators $D_\pm D_\mp$ appearing in~(\ref{3d-spatial}) and
(\ref{4d-spatial-2}).

When $\mu=0$ and we take the zero temperature limit ($\beta\to\infty$)
then the effective action in~(\ref{eff-6}) reduces to
\begin{equation}
  S_{\rm eff}\left|_{\mu=0;\beta\to\infty}\right.= \Tr \left(|{\cal
      E}|\right)
\label{zerotmu}
\end{equation}
which is the familiar $T=0$, $\mu=0$ effective action in a static
magnetic background~\cite{salam}.

When $\mu\neq 0$ and $\beta\to\infty$ we must distinguish between the
cases $\mu<m$ and $\mu>m$. When $\mu<m$, all low temperature thermal
excitations are exponentially suppressed because $|{\cal E}|\pm\mu >0$
(since $|{\cal E}|\geq m$). Therefore, in this case the effective
action reduces just as in (\ref{zerotmu}). However, if $\mu>m$, then
in the infinite $\beta$ limit the first logarithmic term
in~(\ref{eff-6}) contributes for the portion of the spectrum for which
$|{\cal E}|-\mu <0$. Thus, we have
\begin{eqnarray}
  S_{\rm eff}|_{\mu\neq 0;\beta\to\infty} &=& \Tr \biggl[|{\cal E}|+
    \left(\mu-|{\cal E}|\right) \theta{\left(\mu-|{\cal E}|\right)}
    \biggr]\nonumber\\
&=&\Tr \biggl[\mu \; \theta{\left(\mu-|{\cal E}|\right)}
    +|{\cal E}| \; \theta{\left(|{\cal E}|-\mu\right)}  \biggr]
\label{zerot}
\end{eqnarray}
where $\theta$ is the step function. This is the standard expression
for the effective action at zero temperature and with nonzero chemical
potential \cite{shuryak,chodos}. The
first equality in (\ref{zerot}) emphasizes the correction from the zero $T$ and
zero $\mu$ expression (\ref{zerotmu}), while the second emphasizes the
physical content of the effective action with zero $T$ and nonzero $\mu$ as
$\mu$ times the number of occupied particle states plus the trace of the energy
eigenvalues above the threshold $\mu$ \cite{chodos}. In Section VII we examine
this $T=0$ limit in detail for fermions in a static magnetic background.

The form of the effective action~(\ref{eff-6}) is reminiscent of
the grand partition function in non-relativistic statistical mechanics.
Indeed, the non-relativistic limit corresponds to the situation in
which the rest mass energy $m$ is the dominant contribution to $|{\cal E}|$,
\begin{equation}
  |{\cal E}| = \sqrt{m^2+\left(\vec{D}\cdot\vec{\gamma}\right)^2} =
  m+{1\over 2m} \left(\vec{D}\cdot\vec{\gamma}\right)^2+\dots
\label{nr}
\end{equation}
In this limit, with $\mu>m$ and $m\to\infty$, the first logarithmic
term in (\ref{eff-6}) dominates over the second ({\it i.e} antiparticles are
suppressed), and we are left with
\begin{equation}
  S_{\rm eff} \to {1\over \beta} \Tr\,\Log\left(1+e^{\beta
      \mu_{\text{NR}}}e^{-\beta \left(\vec{D}\cdot\vec{\gamma}\right)^2
      /2m}\right)
\label{nreff}
\end{equation}
where we have identified $\mu_{\text{NR}}=\mu-m$ as the non-relativistic
chemical potential. The expression (\ref{nreff}) is $1/\beta$ times the
logarithm of the grand partition function for the corresponding
non-relativistic fermion system.

\section{Proper Time Formulation}
\label{sec:propertime}

The proper time formulation provides an efficient method for computing
the effective action at zero temperature \cite{schwinger,salam,blau}, and
furthermore has the virtue that the generalization to finite
temperature and nonzero chemical potential naturally separates out the
influence of a static background magnetic field. Using an integral
representation of the logarithm to define the logarithm of an
operator, we may express the finite temperature version of the
effective action~(\ref{eff-4}) as
\begin{eqnarray}
  S_{\rm eff}&=&{1\over \beta}\sum_{n=-\infty}^\infty \Tr\,\Log\left[
    -{(\omega_n-\mu)^2\over m^2}+{|{\cal E}|^2\over m^2}\right]
  \nonumber\\ &=& - {1\over
    \beta} \sum_{n=-\infty}^\infty \int_0^\infty {ds\over s} \;
  \Tr \left[ \exp \left( {(\omega_n-\mu)^2\over m^2}s-{|{\cal E}|^2\over
        m^2}s\right) \right]
\label{log}
\end{eqnarray}
where we have chosen to refer all energy scales to $m$ in order to
have dimensionless operators in the exponent. For massless theories one must
choose a different reference energy scale, as discussed in Section VI.

Expression (\ref{log}) may be re-cast in terms of the second elliptic
theta function $\theta_2(u | \tau)$,
\begin{equation}
  S_{\rm eff}=- {1\over \beta} \int_0^\infty {ds\over s} \;
  \theta_2 {\left( {2\pi \mu s\over\beta m^2}\left| {4\pi i s\over
        \beta^2m^2} \right. \right)} \Tr\left[ \exp \left( -{(|{\cal
          E}|^2-\mu^2)\over m^2}s \right) \right]
\label{th}
\end{equation}
where~\cite{bateman,bellman}
\begin{equation}
  \theta_2{\left(u | \tau\right)} = 2 \sum_{n=0}^\infty
  e^{i\pi\tau(n+1/2)^2} \cos \left((2n+1)u\right)
\label{theta2}
\end{equation}
A Poisson summation formula converts the second theta function into a
fourth theta function according to the identity~\cite{bateman}:
\begin{equation}
  \theta_4{\left({u\over \tau}\left|-{1\over
        \tau} \right. \right)} =\left({i\over \tau}\right)^{-1/2}
  e^{iu^2/(\pi \tau)} \; \theta_2 {\left(u | \tau \right)}
\label{poisson}
\end{equation}
This converts the expression~(\ref{th}) for the effective action into
a form involving $\theta_4$:
\begin{equation}
  S_{\rm eff}=- {m\over 2\sqrt{\pi}}\int_0^\infty{ds\over s^{3/2}} \;
  \theta_4{\left({i\beta \mu \over 2}\left| { i \beta^2m^2 \over 4\pi
          s}\right.\right)} \Tr\left[ \exp\left(-{|{\cal E}|^2\over
        m^2}s\right)\right]
\label{th4}
\end{equation}
where the fourth theta function is
\begin{equation}
  \theta_4 {\left(u | \tau\right)} =1+2\sum_{n=1}^\infty (-1)^n
  e^{i\pi\tau n^2} \cos\left(2nu\right)
\label{theta4}
\end{equation}

This expression (\ref{th4}) for the effective action is particulary useful as
it shows clearly the separation between the effects of the static
background magnetic field, which appear solely in the trace factor,
and the effects of finite temperature and nonzero chemical potential,
which appear solely in the $\theta_4\left({i\beta \mu \over 2}\left| {
      i \beta^2m^2 \over 4\pi s}\right)\right.$ factor. Indeed, when $\mu\equiv
0$ and $\beta\to\infty$, the $\theta_4$ factor reduces to $1$ and we are left
with the standard proper time expression for the zero temperature effective
action in a static system. This corresponds to keeping just the term ``$1$'' in
the
expansion (\ref{theta4}) of the $\theta_4$ function, so the remaining summation
over $n$ in (\ref{theta4}) represents the nonzero temperature correction. The
utility of elliptic theta functions in the computation of finite temperature
effective actions has been noted previously in \cite{dittrich} for fermions
(without chemical potential) and in \cite{braden} for bosonic systems.

All information about the static magnetic background is
neatly encapsulated in the proper time propagator
\begin{equation}
  \Tr\left[ \exp\left(-{|{\cal E}|^2\over m^2}s\right)\right] = e^{-s}
  \,\,\Tr\left[ \exp\left(-{(\vec{D}\cdot\vec{\gamma})^2\over
        m^2}s\right)\right]
\label{propagator}
\end{equation}
which is computed independent of any reference to temperature or
chemical potential. Thus, if one computes the zero temperature
effective action using the proper time method it is completely
straightforward to then write down an expression for the effective
action at finite temperature and at nonzero chemical potential simply
by inserting the $\theta_4$ factor as in~(\ref{th4}).

However, while the expression (\ref{th4}) illustrates the separate roles of
finite $\beta$, $\mu$ and the external static field, it is not so
straightforward to use it to obtain useful {\it numerical} estimates. This is
because of the wildly oscillatory behavior of the $\theta_4$-function
in~(\ref{th4}) for large values of the proper time parameter $s$. This
oscillatory behavior is also sensitive to the magnitude of the dimensionless
parameter $\beta\mu$ which appears in the first argument of the $\theta_4$
function. These difficulties are further complicated by the proliferation of
energy scales - for zero $T$ and $\mu$ the only scales are the fermion mass
$m$, the characteristic strength scale $B$
(with dimensions of $m^2$) of the external magnetic field, and
possibly also characteristic length scales associated with spatial
variations in the magnetic field. The generalization to nonzero
temperature and chemical potential introduces two further energy scales:
$\beta$ and $\mu$. In this paper, we concentrate mainly on
high temperature expansions (in which $T=1/\beta$ is the dominant energy
scale), although we also discuss the zero temperature limit (with nonzero
chemical potential) in Section VII.

To conclude this brief review of the finite temperature formalism for fermionic
systems, we show how the general expression~(\ref{th4}) relates to the previous
expression~(\ref{eff-6}), for which our statistical mechanics intuition is most
direct.  Using~(\ref{theta4}) and rescaling the proper time
variable $s$ in~(\ref{th4}), $s/m^2\to\beta^2s/4$, we obtain
\begin{eqnarray}
  S_{\rm eff}&=&- {1\over \beta\sqrt{\pi}}\int_0^\infty{ds\over
    s^{3/2}} \; \theta_4 {\left({i\beta\mu\over 2}\left | {i\over \pi
        s} \right. \right)}  \,\Tr \left[e^{-\beta^2{\cal E}^2
      s/4}\right]\nonumber \\ &=&- {1\over
    \beta\sqrt{\pi}}\int_0^\infty{ds\over
    s^{3/2}}\left[1+2\sum_{n=1}^\infty (-1)^ne^{-n^2/s} \cosh(n\beta
    \mu)\right]\,\Tr \left[e^{-\beta^2{\cal E}^2 s/4}\right]
\label{prelim}
\end{eqnarray}
The integrations over $s$ may now be performed, with those in the
summation term requiring the identity (see \cite{gradshteyn}
Eqs. 3.471.9 and 8.469.3)
\begin{equation}
  {1\over \sqrt{\pi}}\int_0^\infty{ds\over s^{3/2}}\; \exp\left(-{n^2\over
      s}-{\beta^2{\cal E}^2\over 4}s\right)={e^{-\beta n |{\cal
        E}|}\over n}, \qquad n>0
\label{bessel}
\end{equation}
Thus
\begin{equation}
  S_{\rm eff}= \Tr|{\cal E}|+{2\over \beta}\sum_{n=1}^\infty
  {(-1)^{n+1}\over n}e^{-\beta n |{\cal E}|} \cosh(n\beta \mu )
\end{equation}
which is just expression~(\ref{eff-6}).

\section{Free Theories}

Before discussing the temperature expansions for fermions in the
presence of background magnetic fields, we first describe the high
temperature expansions for free theories. This is partly to establish some
notation and to introduce some number theoretic functions (the $zeta$ and $eta$
functions), but also in order to point out some important subtleties that arise
even in the free theories and which have implications for the magnetic systems.

In the free fermionic theory in $(d+1)$-dimensional spacetime, ${\cal
  E}^2=m^2+\vec{p}{\,}^2$, where $\vec{p}$ is a $d$-dimensional momentum
vector.  Thus, the proper-time propagator~(\ref{propagator}) is simply
\begin{equation}
  \Tr\left[ \exp\left(-{{\cal E}^2\over
        m^2}s\right)\right]=\frac{c}{2}\left({m\over
      2\sqrt{\pi}}\right)^d\,{e^{-s}\over s^{d/2}}
\end{equation}
where $c$ is the spinor dimension (the $d=2$ and $d=3$ cases discussed
previously correspond thus to $c=4$). Therefore, the effective
action~(\ref{th4}) is
\begin{equation}
  S_{\rm eff} = - \frac{c}{2}\left({m\over
2\sqrt{\pi}}\right)^{d+1}\int_0^\infty
  {ds\over s^{d+3\over 2}}\,e^{-s}\,\left[1+2\sum_{n=1}^\infty(-1)^n
    \cosh(\beta\mu n) \exp \left(-{\beta^2m^2n^2\over
        4s}\right)\right]
\label{free}
\end{equation}

This expression already illustrates an important difference between
the $3+1$ and $2+1$ dimensional cases, for which $(d+3)/2$ is an
integer or half-odd-integer, respectively. For example, in the zero
temperature and zero chemical potential case, the square parentheses in
(\ref{free}) reduce to a single term ``$1$'', and so the $3+1$ free case
naively
leads to a divergent factor $\Gamma(-2)$ which must be regulated consistently.
This may be achieved, for example, by cutting off the lower limit $0$ of the
$s$
integration, or by shifting the dimension $d\to 3+\epsilon$ and
extracting a finite piece. In contrast, in $2+1$ dimensions the
corresponding factor is $\Gamma(-3/2)$, which is finite. These issues
are well understood for the zero temperature theories
\cite{schwinger,salam,blau}, but below we illustrate analogous differences
between $d$ odd and $d$ even for the $\beta$ and $\mu$ dependent
contribution to the effective action.

The $s$ integrations in the $n$ summation in~(\ref{free}) may be
performed using the following integral representation of the modified
Bessel function (see \cite{bateman}, Eq. 7.12.23)
\begin{equation}
  K_\nu (z)={1\over 2}\left({z\over 2}\right)^\nu \int_0^\infty
  {ds\over s}\, s^{-\nu} \exp\left[-\left(s+{z^2\over
        4s}\right)\right], \qquad {\rm Re} \, z^2 >0, \; |\arg z| <
  \frac{\pi}{2}
\label{mod}
\end{equation}
This leads to\footnote{As mentioned previously, the $1$ in the square bracket
in Eq.~(\ref{free}) corresponds to the zero temperature (and zero chemical
potential) proper time representation of the effective action, which we denote
by $S_{\rm eff}^{T=0;\mu=0}$.}
\begin{equation}
  S_{\rm eff}=S_{\rm eff}^{T=0;\mu=0} -2 c\left({m\over
      2\sqrt{\pi}}\right)^{d+1}\sum_{n=1}^\infty(-1)^n \cosh(\beta\mu
  n) K_{{d+1\over 2}}\left( \beta m n\right) \left({\beta m n\over
      2}\right)^{-(d+1)/2}
\label{free-2}
\end{equation}
For free theories the only energy scales are $m$, $\mu$ and $T=1/\beta$, so a
high temperature expansion corresponds to $\beta m \ll 1$ and
$\beta\mu \ll 1$.
Therefore, expression (\ref{free-2}) may be converted into a high temperature
expansion by using the ascending series expansions of the
modified Bessel function $K_\nu (z)$. The difference between $3+1$ and
$2+1$ dimensions is reflected in the fact that the series expansions
of $K_\nu(z)$ are very different for $\nu$ an integer or $\nu$ a
half-odd-integer.

For $\nu=\text{integer}=N$, the expansion of $K_N(z)$ for small $z$
begins with a $z^{-N}$ term and also involves a logarithmic piece:
\begin{eqnarray}
  K_N(z)&=&{1\over 2}\left( \frac{z}{2} \right)^{-N}
  \sum_{j=0}^{N-1}{(N-j-1)!\over j!}\left(-{z^2\over 4}\right)^j
  \nonumber \\ &&-(-1)^N \left( \frac{z}{2} \right)^N
  \sum_{j=0}^\infty {(z^2 / 4)^j\over j!(N+j)!}\left[ \log{z\over 2}-
    \frac{1}{2} \psi(j+1) - \frac{1}{2} \psi(N+j+1)\right]
\label{integer}
\end{eqnarray}
where $\psi(x)={d\over dx} \log \Gamma(x)$ is the digamma function
\cite{bateman}.  On the other hand, for $\nu= \text{half-odd-integer}
= N+{1\over 2}$, the corresponding small $z$ expansion is
\begin{eqnarray}
  K_{N+{1\over 2}}(z) &=& \sqrt{{\pi\over
      2}}z^{-N-1/2}\,e^{-z}\,\sum_{j=0}^N \frac{(N+j)!}{j!(N-j)!}
  \frac{z^{N-j}}{2^j}\nonumber\\
  &=& {1\over 2} \sum_{j=0}^\infty \left[
    {\Gamma(N-j+\frac{1}{2})\over j!} \; \left({z\over2}\right)^{-N-1/2} +
    {\Gamma(-N-j-\frac{1}{2})\over j!} \; \left({z\over2}\right)^{N+1/2}
  \right] \left(-{z^2\over 4}\right)^j
\label{half}
\end{eqnarray}
which begins with $z^{-N-1/2}$, but has no logarithmic piece.

These Bessel function expansions~(\ref{integer}) and~(\ref{half}) lead to the
following explicit high temperature expansions of the free
effective action (\ref{free-2}). In $2+1$ dimensions, or
more generally in $d+1$ dimensions with $d$ even,
\begin{eqnarray}
  S_{\rm eff}^{d\text{\ even}} &=& S_{\rm eff}^{T=0;\mu=0}
  -c\left(
    \frac{m}{2\sqrt{\pi}} \right)^{d+1} \sum_{n=1}^\infty (-1)^n
  \cosh(\beta\mu n) \nonumber\\
  && \hspace{1cm} \times \sum_{j=0}^\infty \left[ \frac{\Gamma({d+1\over
        2}-j)}{j!} \left({\beta m n\over 2}\right)^{-(d+1)} +
    \frac{\Gamma(-(\frac{d+1}{2})-j)}{j!} \right]
  \left(-{\beta^2m^2n^2\over 4}\right)^j
\label{free-3}
\end{eqnarray}
Notice that the $\cosh(\beta \mu n)$ term can be expanded in powers of
$n^2$, in which case the summation over $n$ has the form of the eta
function \cite{bateman} (an alternating series analogue of the Riemann
zeta function)
\begin{equation}
  \eta(r)=\sum_{n=1}^\infty(-1)^{n-1} {1\over n^r}
\end{equation}
For $\text{Re} \, r > 1$ the eta function $\eta(r)$ is related to the Riemann
zeta function, $\zeta(r)=\sum_{n=1}^\infty 1/n^r$, by
\begin{equation}
  \eta(r)=(1-2^{1-r})\zeta(r)
\end{equation}
For negative integer values of $r$, this function takes especially simple
values.
For even negative integers
\begin{equation}
  \eta(0)={1\over 2} ,\hskip 2cm \eta(-2k)=0 \qquad k=1,2,3,\ldots
\end{equation}
while for odd-integer negative arguments
\begin{equation}
  \eta(1-2k)=(2^{2k}-1) {{\cal B}_{2k}\over 2k} \qquad k=1,2,3,\ldots
\end{equation}
where ${\cal B}_{2k}$ is the $(2k)^{th}$ Bernoulli number. Also note
that $\eta(1)= \log 2$.

Since $\eta(-2j)$ vanishes for $j=1,2,\dots$, only the $j=0$ term survives in
the second summation term inside the
square brackets in~(\ref{free-3}), and this surviving term then cancels against
the zero $T$, zero $\mu$ effective action. In contrast, the first summation
term inside the square brackets in~(\ref{free-3}) involves odd powers of $n$,
and so contributes for all $j$. Finally, we obtain the
following double expansion
\begin{eqnarray}
  S_{\rm eff}^{d\text{\ even}}&=&
  c\left(\beta\sqrt{\pi}\right)^{-(d+1)} \sum_{k=0}^\infty {
    (\beta\mu)^{2k}\over (2k)!}\sum_{j=0}^\infty {(-1)^j
    \Gamma({d+1\over 2}-j)\eta(d+1-2k-2j)\over j!}\left({\beta m\over
      2}\right)^{2j}
\label{free-4}
\end{eqnarray}
which agrees with the corresponding expansion in Eq. (4.24)
of Ref. \cite{actor}, although the sum is organized in a different manner.
Notice that the leading term at high temperature is proportional to
$T^{d+1}$, with the remaining terms in the expansion involving higher
powers of $\beta m$ or $\beta\mu$. For zero chemical potential the expansion
simplifies to a single sum
\begin{eqnarray}
  S_{\rm eff}^{d\text{\ even}}|_{\mu=0} &=&
  c\left(\beta\sqrt{\pi}\right)^{-(d+1)} \sum_{j=0}^\infty {(-1)^j
    \Gamma({d+1\over 2}-j)\eta(d+1-2j)\over j!}\left({\beta m\over
      2}\right)^{2j}
\label{free-5}
\end{eqnarray}
and in the massless case
\begin{eqnarray}
  S_{\rm eff}^{d\text{\ even}}|_{m=0} &=&
  c\left(\beta\sqrt{\pi}\right)^{-(d+1)}
\Gamma(\case{d+1}{2})\sum_{k=0}^\infty
{(\beta\mu)^{2k}\eta(d+1-2k)\over (2k)!}
\end{eqnarray}

In $3+1$ dimensions, or more generally for $d+1$ dimensions with $d$
odd, the Bessel function expansion~(\ref{integer}) leads to the
following high temperature expansion of the free effective action:
\begin{eqnarray}
  S_{\rm eff}^{d\text{\ odd}}&=&S_{\rm eff}^{T=0;\mu=0} -
  2c\left({m\over 2\sqrt{\pi}}\right)^{d+1} \sum_{n=1}^\infty(-1)^n
  \cosh(\beta\mu n) \nonumber\\ &&\hspace{.2cm} \times \left[{1\over
      2} \left({\beta m n\over 2}\right)^{-(d+1)}
    \sum_{j=0}^{\frac{d-1}{2}} {\Gamma({d+1\over 2}-j)\over
      j!}\left(-{\beta^2m^2n^2\over 4}\right)^j \right. \nonumber\\
  &&\hspace{.4cm} \left. - (-1)^{(d+1)/2} \sum_{j=0}^\infty
    {(\beta^2m^2n^2/ 4)^j \over j! (j+(d+1)/2)!}\left(\log\case{\beta
        m n}{2}- \frac{1}{2} \psi(j+1)- \frac{1}{2}
      \psi(j+\case{d+3}{2})\right)\right]
\label{free-6}
\end{eqnarray}
The first term inside the square brackets produces $T^{d+1}$ times a
polynomial $P_{d+1}(\beta\mu,\beta m)$ of order $d+1$ in $\beta\mu$
and $\beta m$, because the expansion involves only even powers of $n$,
for which the eta function vanishes if these are positive even powers,
but gives a finite contribution while the power is negative or zero.
Thus the summation over $j$ and the expansion of $\cosh(\beta \mu n)$
are each truncated. The effective action takes then the form
\begin{eqnarray}
  S_{\rm eff}^{d\text{\ odd}} &=& S_{\rm eff}^{T=0;\mu=0} - c \left(
    \frac{m}{2\sqrt{\pi}} \right)^{d+1} \Biggl\{ (\beta m)^{-(d+1)}
  P_{d+1}(\beta\mu,\beta m) \nonumber\\
  && \qquad +
  \frac{(-1)^{(d+1)/2}}{\Gamma((d+3)/2)} \left[ \log\case{\beta
      m}{4\pi} - \frac{1}{2} \psi(1)- \frac{1}{2} \psi(\case{d+3}{2})
  \right] + {\rm Tayl}(\beta m, \beta \mu) \Biggr\}
\end{eqnarray}
where the Taylor series ${\rm Tayl}(\beta m, \beta \mu)$ is determined
using the following further summation identity (for details see
\cite{actor}):
\begin{equation}
  \sum_{n=1}^\infty (-1)^n n^{2k} \log(2\pi n)=\cases{-\log2, \hskip
    1cm k=0\cr \displaystyle (-1)^k {1\over 2} (1-2^{2k+1})
    \frac{(2k)!}{(2\pi)^{2k}}\zeta(2k+1), \hskip 1cm k\in Z^+}
\end{equation}

\section{Constant Magnetic Background}

As discussed in Section~\ref{sec:propertime}, in order to obtain a closed
expression of the form (\ref{th4}) for the
finite temperature and finite chemical potential effective action for
fermions in a static background magnetic field one simply needs the proper
time propagator~(\ref{propagator}).  For the case of a uniform
background field this is a well-known computation whose exact
solubility is due to the fact that the spectra of the operators
$-D_\pm D_\mp$ correspond to one-dimensional harmonic oscillators, with
`frequency' $2B$,
each energy level having degeneracy given by the (infinite) background
magnetic flux \cite{landau}.
\begin{eqnarray}
  -D_+D_-&=&2Bn\hskip 2cm n=1,2,3,\dots \nonumber\\
  -D_-D_+&=&2Bn\hskip 2cm n=0,1,2,3,\dots
\label{landau}
\end{eqnarray}
Thus, the proper-time propagator is
\begin{equation}
  \Tr\left[\exp\left(-{{\cal E}^2\over m^2}s\right)\right]=
  \left\{ \begin{array}{lr} {B\over 2\pi}\left({m\over
          2\sqrt{\pi}}\right) \frac{1}{\sqrt{s}}\,e^{-s}\, \coth\left({B
          s\over m^2}\right), & \qquad 3+1\,\, \text{dim.}\\ {B\over
        2\pi}e^{-s}\, \coth\left({B s\over m^2}\right), & 2+1\,\,
      \text{dim.} \end{array} \right.
\end{equation}
The extra $\left({m\over 2\sqrt{\pi}}\right) {1\over \sqrt{s}}$ factor
in $3+1$ dimensions is due to the integration over $k_3$, the momentum
corresponding to the free motion in the $z$ direction and $B/2\pi$ is
the usual Landau level degeneracy factor \cite{landau}.

The effective action~(\ref{th4}) is therefore
\begin{equation}
  S_{\rm eff} = - \left({m\over 2\sqrt{\pi}}\right)^{d-1} {B\over
    2\pi} \int_0^\infty {ds\over s^{(d+1)/2}}\, e^{-s}\,
  \left[\coth\left({Bs\over m^2}\right)-{m^2\over Bs} \right]
  \theta_4{\left({i\beta\mu\over 2}\left|
      {i\beta^2m^2\over 4\pi s} \right. \right)}
\label{mag}
\end{equation}
where $d=2$ or $d=3$. Note that we have explicitly subtracted the zero $B$
contribution by subtracting the leading small $s$ divergence $m^2/(Bs)$ from
the
function $\coth(Bs/m^2)$. This ensures that the effective
action vanishes with vanishing $B$, and simply corresponds to computing a
difference of the effective action with and without the $B$ field. We could
also subtract the $B^2$ contribution as this
may be absorbed into the original action via a charge renormalization. However,
we choose to defer this additional subtraction for two reasons.
First, we wish to emphasize that the $B^2$ contribution differs in an
interesting manner for $d=2$ and $d=3$. Second, it is conventional
\cite{schwinger,salam,blau} to make this subtraction only for the zero $T$ and
zero
$\mu$ portion of the effective action, since in the $3+1$ dimensional case this
contribution involves a divergence whose regularization involves a charge
renormalization.

To develop high temperature expansions from (\ref{mag}), along the lines
discussed for the free case in the previous Section, would require expanding
the $\theta_4$ factor in the integrand and then performing the proper time
integrals. However, these integrals cannot be done in closed form (in contrast
to the free case where they led to modified Bessel functions - see (\ref{mod}))
because of the $\coth(Bs/m^2)$ factor. We therefore rescale
the proper time parameter $s$: $s/m^2 \to \beta^2 s$. Then in the high
$T$ limit we expand the $\coth(\beta^2 B s)$ factor:
\begin{equation}
  \coth\left({\beta^2Bs}\right)-{1\over \beta^2Bs}
  =\sum_{k=1}^\infty{2^{2k}{\cal B}_{2k}\over (2k)!} \left(\beta^2 B s
  \right)^{2k-1}
\end{equation}
where ${\cal B}_{2k}$ are the Bernoulli numbers \cite{bateman}. This
leads to the following double expansion
\begin{eqnarray}
  S_{\rm eff} &=& S_{\rm eff}^{T=0;\mu=0}- 8 \left( \beta \sqrt{2\pi}
  \right)^{-(d+1)}  \sum_{k=1}^\infty
  {{\cal B}_{2k}\left({\beta^2B}\right)^{2k} \over (2k)!}
  \sum_{n=1}^\infty (-1)^n {\cosh(\beta\mu n)\over n^{d+1-4k}} {K_{2k-{d+1\over
      2}}\left(\beta mn\right) \over (\beta mn)^{2k-{d+1\over 2}}}
\label{s-bess}
\end{eqnarray}
Assuming that the temperature $T$ dominates all other energy scales, we can
further expand using the series representations~(\ref{integer})
and~(\ref{half}) of the modified Bessel functions $K_\nu(z)$.  Note that, just
as in the free case, there is a significant difference between $d=2$ and $d=3$,
as the former case involves Bessel functions with half-odd-integer index, while
the latter involves Bessel functions with integer index.

For $d=2$,
\begin{eqnarray}
  S_{\rm eff}^{d=2} &=& {4 \over\left( \beta\sqrt{\pi} \right)^{3}}
  \sum_{k=1}^\infty{{\cal B}_{2k}({\beta^2 B\over 2})^{2k}\over (2k)!}
  \sum_{j=0}^\infty {\left(-{\beta^2m^2\over 4}\right)^j
    \Gamma\left(\frac{3}{2}-j-2k\right)\over j!} \sum_{l=0}^\infty
  {(\beta \mu)^{2l} \eta(3-2j-4k-2l)\over (2l)!}
\label{dc1}
\end{eqnarray}
Notice that the zero $T$, zero $\mu$ effective action is canceled by a term
arising from the expansion of the Bessel functions. Considered as an expansion
in the magnetic field, the leading term is of the form
\begin{equation}
  S_{\rm eff}^{d=2} = \beta B^2 f(\beta\mu,\beta m) + \cdots
\end{equation}
where $f(\beta\mu,\beta m)$ is a dimensionless function whose leading
term is a (negative) constant. The expression (\ref{dc1}) has a
well-defined massless limit:
\begin{equation}
  S_{\rm eff}^{d=2}|_{m=0} = {4 \over\left(\beta \sqrt{\pi}
  \right)^{3}} \sum_{k=1}^\infty{{\cal B}_{2k}(\beta^2 B/ 2)^{2k}
    \Gamma\left(\frac{3}{2}-2k\right)\over (2k)!} \sum_{l=0}^\infty
  {(\beta \mu)^{2l} \eta(3-4k-2l)\over (2l)!}
\label{....}
\end{equation}
And for zero chemical potential,
\begin{equation}
  S_{\rm eff}^{d=2}|_{\mu=0}= {4\over \left( \beta \sqrt{\pi} \right)^{3}}
  \sum_{k=1}^\infty{{\cal B}_{2k}(\beta^2 B / 2)^{2k}\over (2k)!}
  \sum_{j=0}^\infty {\left(- {\beta^2m^2\over 4}\right)^j
    \Gamma\left(\frac{3}{2}-j-2k\right)\eta(3-2j-4k)\over j!}
\label{zeromu}
\end{equation}

For $d=3$ the situation is somewhat different owing to the appearance of a
divergent term in the zero $T$, zero $\mu$ effective action and of logarithmic
terms in both the zero $T$, zero $\mu$ piece and in the finite temperature
piece of the effective action. From (\ref{s-bess}), using the Bessel function
expansion (\ref{integer}), we find
\begin{eqnarray}
 S_{\rm eff}^{d=3} &=& -\frac{B^2}{12\pi^2} \left( \log \frac{\beta
      m}{\pi} + 6 \pi^2 C - \psi(1) \right) + \frac{64 \pi^2}{\beta^4}
  \sum_{k=1}^\infty \frac{{\cal B}_{2k}}{(2k)!} \left(
    \frac{\beta^2 B}{8\pi^2} \right)^{2k} \sum_{j=0}^\infty
  \frac{\left( - \frac{\beta^2m^2}{16\pi^2} \right)^j}{j!}
  \sum_{l=0}^\infty \frac{\left( - \frac{\beta^2\mu^2}{4\pi^2}
    \right)^l}{(2l)!} \nonumber\\
 && \qquad \times \left( 1 - 2^{4k+2j+2l-3} \right)
  \zeta(4k+2j+2l-3)
  \frac{\Gamma(4k+2j+2l-3)}{\Gamma(2k+j-1)}\Biggr|_{\text{except the term}
    \atop k=1\; j=0\; l=0}
\label{three}
\end{eqnarray}
Notice that the  $T=0$, $\mu=0$ effective action cancels against a
term
\begin{equation}
{m^4\over 8\pi^2}\sum_{k=2}^\infty { {\cal B}_{2k}
\left(2B/m^2\right)^{2k}(2k-3)!\over (2k)!}
\end{equation}
arising from the Bessel function expansions, leaving a term
quadratic in the $B$ field with an (infinite) coefficient
\begin{equation}
  C = \frac{1}{12\pi^2} \int_0^\infty ds s^{-1} \, e^{- m^2 s}
\end{equation}
which may be absorbed in a renormalization of the electric
charge~\cite{schwinger}.

The remaining expansion (\ref{three}) has no quartic or
quadratic term in the temperature but has a logarithmic dependence on
the temperature. This agrees with Ref.~\cite{elmfors} but disagrees
with Ref.~\cite{dittrich}.

To conclude this discussion of the high temperature limit for fermions
in a constant magnetic background, we observe that in the $2+1$
dimensional case it is possible to give an alternative derivation of
the magnetic background effective action beginning instead with the
expression~(\ref{eff-5}). For ease of presentation, consider the case
$\mu=0$. Then
\begin{equation}
  S_{\rm eff} = {2\over \beta}\Tr\,\Log\, \cosh{\beta\over 2}|{\cal
    E}| = {2\over \beta}\sum_{p=1}^\infty {(2^{2p}-1){\cal
      B}_{2p}\over (2p)(2p)!} \beta^{2p}\Tr\left[ |{\cal
      E}|^{2p}\right]
\label{direct}
\end{equation}
The trace over the even powers of the energy operator can be expressed
in terms of generalized zeta functions. In $2+1$
dimensions
\begin{equation}
  \Tr\left[ |{\cal E}|^{2p}\right] = \frac{B}{2\pi} \left[ 2
    \sum_{n=1}^\infty (m^2+2Bn)^p + m^{2p} \right] =
  \frac{(2B)^{p+1}}{2\pi} \zeta{\left(-p,1+\frac{m^2}{2B}\right)} +
  \frac{B}{2\pi} m^{2p}
\end{equation}
where the generalized zeta function is defined as \cite{bateman}
\begin{equation}
  \zeta(s,q)=\sum_{n=0}^\infty {1\over (n+q)^s}
\end{equation}
When the first argument of the generalized zeta function is a negative integer,
it reduces to a polynomial (in the second argument) which is in fact
proportional to a Bernoulli polynomial:
\begin{equation}
\zeta(-p,x)= -{{\cal B}_{p+1}(x)\over p+1}
\end{equation}
The Bernoulli polynomials are given by
\begin{equation}
{\cal B}_p(x)=\sum_{r=0}^p\left(\matrix{p\cr r}\right){\cal B}_rx^{p-r}
\end{equation}
with expansion coefficients involving the Bernoulli numbers ${\cal B}_r$.

Thus, the high temperature expansion (\ref{direct}) may be written as
\begin{equation}
S_{\rm eff}^{d=2}|_{\mu=0}=-{1\over 2\pi \beta^3}\sum_{p=1}^\infty
{(2^{2p}-1){\cal B}_{2p} \over
p(p+1)(2p)!}\left(2B\beta^2\right)^{p+1}\left[{\cal B}_{p+1}\left({m^2\over
2B}\right)+\frac{p+1}{2}\left(\frac{m^2}{2B}\right)^p\right]
\label{simple}
\end{equation}
Remarkably, this expansion (\ref{simple}) is simply another way of expressing
the double expansion in (\ref{zeromu}), up to a $B$ independent term.

\section{Zero temperature limit}

Having considered the high temperature behavior of the effective action,
we now turn to the other extreme: $T=0$. With zero chemical
potential this is just the usual field theoretic effective action,
which may be computed using the proper-time
formalism~\cite{schwinger,salam,blau}. With nonzero chemical
potential, the zero $T$ limit is more subtle due to nonanalytic
contributions from the sharp cut-off of the energy spectrum  which is
no longer smoothed out by thermal fluctuations. The low $T$ effective
action with nonzero chemical potential and  background magnetic field
in $3+1$ dimensions has been studied recently~\cite{elmfors,zeitlin},
primarily using the real-time formalism. Here we discuss a somewhat
different (and more direct) approach to the low temperature limit, based
on the expression~(\ref{zerot}). According to~(\ref{zerot}), the
correction $\triangle S$ to the zero $T$ and zero $\mu$ effective
action when the chemical potential is nonzero may be written as
\begin{equation}
  \triangle S|_{T=0;\mu\not=0} = \Tr \biggl[ (\mu - |{\cal E}|) \theta
  {(\mu - |{\cal E}|)} \biggr]
\label{corr}
\end{equation}
Since $|{\cal E}| \geq m$ it is clear that this correction to the usual
zero temperature effective action vanishes unless $\mu > m$. For the
remainder of this Section we assume that $\mu$ is indeed greater than
$m$.

The free case is well known~(see for example \cite{chodos}) and we
turn directly to the case of a magnetic background. We consider the
$2+1$ and $3+1$ dimensional cases separately. In $2+1$ dimensions, to
compute the correction term~(\ref{corr}) we simply fill the Landau
levels, according to the Landau level spectrum (\ref{landau}), up to just below
the level of the chemical potential
\begin{eqnarray}
  \triangle S^{d=2}|_{T=0;\mu\not=0} &=& \frac{B}{2\pi} \left\{
    2 \sum_{n=1}^{\left[ \frac{\mu^2- m^2}{2 B} \right]} (\mu -
    \sqrt{m^2 + 2 B n}) + (\mu - m) \right\} \nonumber\\
  &=& \frac{Bm}{\pi} \Biggl\{ \frac{\mu}{m} \left[ \frac{\mu^2- m^2}{2
        B} \right] + \sqrt{\frac{2B}{m^2}} \; \zeta{\left( - \frac{1}{2},
        1 + \frac{m^2}{2B} + \left[ \frac{\mu^2- m^2}{2 B} \right]
      \right)} \nonumber\\
  && \qquad\quad + \frac{\mu - m}{2 m} - \sqrt{\frac{2B}{m^2}} \;
  \zeta{\left( - \frac{1}{2}, 1 + \frac{m^2}{2B} \right)} \Biggr\}
\end{eqnarray}
The last two terms are in fact canceled by the zero
chemical potential part of the effective action. We note the appearance
of a nonanalytic dependence on the integer part $\left[ \frac{\mu^2-
    m^2}{2 B} \right]$ of the dimensionless ratio $\frac{\mu^2-
  m^2}{2 B}$. In the nonrelativistic limit, this leads to oscillatory
behavior of the effective action and related thermodynamics quantities
(such as the magnetization) as the strength $B$ of the external
magnetic field is varied.

Specifically, in the nonrelativistic, weak field limit, $m^2 \gg (\mu^2 -
m^2) \gg B$, the $2+1$ dimensional correction term is
\begin{equation}
  \triangle S^{d=2}_{\text{NR}}|_{T=0;\mu\not=0} = \frac{B}{2 \pi} \left\{
    2 \sum_{n=1}^{\left[ \frac{m \mu_{\text{NR}}}{B} \right]}
    \left(\mu_{\text{NR}} - \frac{B}{m} n \right) + \mu_{\text{NR}} \right\}
\end{equation}
where $\mu_{\text{NR}} = \mu - m$ is the nonrelativistic chemical
potential. This may be re-expressed as
\begin{mathletters}
\begin{eqnarray}
\lefteqn{
  \triangle S^{d=2}_{\text{NR}}|_{T=0;\mu\not=0} = \frac{B^2}{\pi m} \left( 1 +
    \left[ \frac{m \mu_{\text{NR}}}{B} \right] \right) \left( \frac{m
      \mu_{\text{NR}}}{B} - \frac{1}{2} \left[ \frac{m
        \mu_{\text{NR}}}{B} \right] \right) - \frac{B
    \mu_{\text{NR}}}{2 \pi} } \qquad \label{zero2+1}\\
\nonumber\\[-12pt]
  && = \frac{B^2}{\pi m} \left\{ \zeta{\left( - 1, \frac{m
          \mu_{\text{NR}}}{B} - \left[ \frac{m \mu_{\text{NR}}}{B}
        \right] \right)} - \zeta{\left( - 1, 1+ \frac{m
          \mu_{\text{NR}}}{B} \right)} \right\} - \frac{B
    \mu_{\text{NR}}}{2 \pi} \nonumber\\
  && = \frac{B^2}{4 \pi^3 m} \sum_{k=1}^\infty \frac{\sin \left( 2 \pi
      k \left( \frac{m \mu_{\text{NR}}}{B} - \left[ \frac{m
            \mu_{\text{NR}}}{B} \right] \right) - \frac{\pi}{2}
    \right)}{k^2} - \frac{B^2}{\pi m} \zeta{\left( - 1, 1+ \frac{m
          \mu_{\text{NR}}}{B} \right)} - \frac{B
      \mu_{\text{NR}}}{2\pi}
\end{eqnarray}
\end{mathletters}
where we have used the zeta function identity (see~\cite{bateman} 1.10
Eq. (6))
\begin{equation}
  \zeta(s, v) = 2 (2\pi)^{s-1} \Gamma(1 - s) \sum_{k=1}^\infty k^{s-1}
  \sin (2\pi k v + \pi s / 2), \qquad 0 < v \leq 1, {\rm Re} \, s < 0
\label{zetaosc}
\end{equation}
Thus the effective action includes a term which as a function of $1/B$
is periodic with period $1/(m\mu_{\text{NR}})$, the characteristic
period of the de Haas - van Alphen oscillations~\cite{pathria}. It is
interesting to notice that (in contrast to 3+1 dimensions) there is an
explicit algebraic expression~(\ref{zero2+1}) for the effective action
that shows these oscillations.

In $3+1$ dimensions, the relativistic correction term~(\ref{corr}) is
\begin{eqnarray}
\lefteqn{
  \triangle S^{d=3}|_{T=0;\mu\not=0} = \frac{B}{2\pi} \Biggl\{
    2 \sum_{n=1}^{\left[ \frac{\mu^2- m^2}{2 B} \right]} 2
    \int_0^{\sqrt{\mu^2 - m^2 - 2 B n}} \frac{dp}{2\pi} \; (\mu -
    \sqrt{p^2 + m^2 + 2 B n}) } \qquad \nonumber\\
  &&\qquad\qquad\qquad {} + 2 \int_0^{\sqrt{\mu^2 - m^2 - 2 B n}}
    \frac{dp}{2\pi} \; (\mu - \sqrt{p^2 + m^2}) \Biggr\} \nonumber\\
  &&{} = \frac{B}{(2\pi)^2} \Biggl\{ \mu \sqrt{\mu^2 - m^2} - m^2 \Log
  \left( \frac{\mu + \sqrt{\mu^2 - m^2}}{m} \right) \nonumber\\
  &&{}  + 2 \sum_{n=1}^{\left[ \frac{\mu^2- m^2}{2 B} \right]} \left( \mu
    \sqrt{\mu^2 - m^2 - 2 B n} - (m^2 + 2 B n) \Log \left( \frac{\mu +
        \sqrt{\mu^2 - m^2 - 2 B n}}{\sqrt{m^2 + 2 B n}} \right)
  \right) \Biggr\}
\end{eqnarray}
This expression agrees with a corresponding one in~\cite{zeitlin},
derived from a zero temperature limit in the real-time
formalism. Expanding the logarithmic term in the weak field limit (in
which $B \ll \mu^2$) we obtain
\begin{eqnarray}
\lefteqn{
  \triangle S^{d=3}|_{T=0;\mu\not=0} = \frac{B m^2}{(2\pi)^2} \left\{
    \frac{\mu}{m} \sqrt{\frac{\mu^2 - m^2}{m^2}} - \Log \left(
      \frac{\mu + \sqrt{\mu^2 - m^2}}{m} \right) \right\} }\quad
  \nonumber\\
  &&+ \frac{2
    B^2}{\pi^2} \sum_{l=0}^\infty \frac{\left( \frac{2 B}{\mu^2}
    \right)^{l + 1/2}}{(2 l + 1) (2 l + 3)} \left\{ \zeta{\left( - l -
        \frac{3}{2}, \frac{\mu^2 - m^2}{2 B} - \left[ \frac{\mu^2 -
            m^2}{2 B} \right] \right)} - \zeta{\left( - l -
        \frac{3}{2}, \frac{\mu^2 - m^2}{2 B} \right)} \right\}
\end{eqnarray}
This expression involves a term depending on the nonanalytic
fractional part of $\frac{\mu^2 - m^2}{2 B}$. Again oscillations arise
when we represent the first zeta function with the help
of the zeta function identity~(\ref{zetaosc})
\begin{equation}
  \zeta{\left( - l - \case{3}{2}, \case{\mu^2 - m^2}{2 B} - \left[
        \case{\mu^2 - m^2}{2 B} \right] \right)} = 2 \Gamma(l + 5/2)
  \sum_{k=1}^\infty \frac{\sin \left( 2 \pi k \left(
      \frac{\mu^2 - m^2}{2 B} - \left[ \frac{\mu^2 - m^2}{2 B} \right]
    \right) - \frac{\pi l}{2} - \frac{3\pi}{4} \right)}{(2 \pi k)^{l +
    5/2}}
\end{equation}
These oscillatory components reveal themselves in a nonrelativistic
limit $m^2 \gg (\mu^2 - m^2) \gg B$ where only the first term $l=0$
survives, being responsible for the usual nonrelativistic de
Haas - van Alphen effect~\cite{pathria}:
\begin{equation}
  \Delta S_{\text{NR}}^{d=3}|^{\rm oscill}_{T=0;\mu\neq 0}={B^{5/2}\over
    4m\pi^4}\sum_{k=1}^\infty {\sin\left(2\pi k \frac{m\mu_{NR}}{B}
      -\frac{3\pi}{4}\right)\over k^{5/2}}
\end{equation}

\section{Conclusions}

In this paper we have shown how to compute finite temperature expansions of the
effective action for fermions in the presence of a static background magnetic
field. The computational techniques of the imaginary time formalism (for the
finite temperature effects) and the proper time method (for the background
field effects) combine elegantly to produce a simple formal expression
(\ref{th4}) for the finite $T$, and nonzero chemical potential, effective
action. Explicit high (and low) temperature expansions may then be obtained
given the explicit functional form of the proper time propagator
(\ref{propagator}). Generically, high temperature expansions involve
logarithmic terms in $d+1$ dimensions with $d$ odd, with these terms being
absent if $d$ is even. These logarithmic terms may be traced to corresponding
terms appearing in the series expansions of the modified Bessel function
$K_N(z)$. For a uniform magnetic field background, the zero temperature limit
(with nonzero chemical potential) may also be evaluated in a straightforward
manner in terms of generalized zeta functions. These zero temperature effective
actions
involve oscillatory pieces which reflect the Landau level structure of the
underlying single particle spectrum.

The formalism developed here is motivated in part by the fact that the proper
time method provides a powerful technique for analyzing non-constant background
fields. For example, one could generalize from a {\it uniform} static
background magnetic field to a static magnetic field with spatial dependence.
Such a generalization has been considered recently for the zero temperature
(and zero chemical potential) effective action in $2+1$ dimensions, with one
particular nonuniform background field leading to an exact closed-form
expression for the
effective action \cite{cangemi}. It would be interesting to incorporate the
finite temperature and finite chemical potential effects into such a model.
Since expression (\ref{th4}) completely separates the finite temperature (and
nonzero chemical potential) effects from those of the static background, it is
very tempting to make high (and low) temperature expansions at this level,
without having an explicit expression for the proper time propagator. However,
even in the free case this leads to subtleties which can be missed if one is
not careful. For example, even in the free case, one might think to make a high
temperature expansion by expanding the exponential term inside the proper time
integral in the effective action (\ref{free}). This amounts to deriving the
series expansions (\ref{integer}) and (\ref{half}) for the modified Bessel
function $K_\nu(z)$ directly from the integral representation (\ref{mod}) by
expanding in powers of $z^2$ {\it inside} the $s$ integral. But this type of
expansion does not produce
the necessary logarithmic terms when $\nu$ is an integer - instead, to obtain
the correct series expansion one must use
a different integral representation based on a contour integral \cite{bessel}.
For fermions in the presence of some general background gauge field, such naive
expansions, made {\it inside} the proper time integral, must also be treated
with care, particularly in $3+1$ dimensions.

\vskip 4cm
This work is supported in part by the NSF under contract
PHY-92-18990 and by the DOE under grant DE-FG02-92ER40716.00. We thank the
Aspen Center for Physics for hospitality during the initial stages of this
work.

\end{document}